\documentclass{aastex63}

\usepackage{multirow}
\usepackage{tablefootnote}
\graphicspath{ {images/} }
\usepackage{color,soul}
\submitjournal{ApJ}

\begin{document}
\title{High-Cadence Dispersed Spectral Analysis of Supernova Remnant 1987A}

\correspondingauthor{Evan Bray}
\email{Bray.EvanP@gmail.com}

\author{Evan Bray}
\affiliation{Department of Astronomy and Astrophysics \\
The Pennsylvania State University \\
525 Davey Lab, University Park, PA 16802 USA}

\author{David N. Burrows}
\affiliation{Department of Astronomy and Astrophysics \\
The Pennsylvania State University \\
525 Davey Lab, University Park, PA 16802 USA}

\author{Sangwook Park}
\affiliation{Department of  Physics \\
University of Texas at Arlington \\
Arlington, TX 76019 USA}

\author{Aravind P. Ravi}
\affiliation{Department of  Physics \\
University of Texas at Arlington \\
Arlington, TX 76019 USA}

\begin{abstract}

We present an analysis of the dispersed spectral data from 11 epochs (March 2011 to September 2018) of supernova remnant (SNR) 1987A observations performed with \textit{Chandra}. These observations were performed with the High Energy Transmission Grating (HETG) as part of our ongoing \textit{Chandra} monitoring campaign of SNR 1987A, whose 1$^{st}$-order dispersed spectrum provides a significantly greater energy resolution than the previously-published 0$^{th}$-order spectrum. Our data sets with moderate exposure times of $\sim$50-70ks per epoch cover the time period between deep Chandra HETG observations (with individual exposures $>\sim$200ks) taken in March 2011 and March 2018. These data have a much higher cadence than the widely-spaced deep high-resolution spectra, at the expense of total exposure time. While statistical uncertainties are large due to low photon count statistics in the observed 1$^{st}$-order spectra, we find that spectral model parameters are generally in line with the shock wave propagating into the medium beyond the dense inner ring, as suggested by \cite{Frank2016}. As the reverse shock begins ionizing the heavier elements of the supernova ejecta interior to the equatorial ring, spectral fit parameters are expected to change as the chemical makeup and physical properties of the shocked gas evolve. Based on our broadband spectral model fits, we find that abundance values appear to be constant in this time period. While our results are somewhat limited due to photon statistics, we demonstrate the utility of the dispersed HETG spectral analysis that can be performed with our regular Chandra monitoring observations of SNR 1987A.

\end{abstract}

\section{Introduction} \label{sec:introduction}
As the only supernova to occur in the immediate vicinity of the Milky Way in recent times, SN 1987A has been the subject of many studies in the past 33 years. Observational campaigns at various wavelengths have been conducted to study the formation and evolutionary characteristics of the resulting supernova remnant (SNR). Ground-based observations of the remnant by \cite{Crotts1989} and \cite{Wampler1990} indicated the existence of at least two outer rings that had been flash-ionized by the explosion itself. After the launch of the \textit{Hubble Space Telescope}, it was revealed that the system is comprised of three ionized rings; one in the equatorial plane and two larger ones above and below it (\cite{Burrows1995}). Around the same time, \textit{ROSAT} generated the first X-ray light curve for the remnant, and correctly predicted the sudden brightening phase that began in the mid-2000's (\cite{Chevalier1995,Hasinger1996}). This era of increasing flux lasted for about a decade, and was caused by the SN blast wave reaching the inner edge of the equatorial ring (ER) and shock-heating it to X-ray temperatures. After the launch of \textit{Chandra}, bright lobes of X-ray emission were identified by utilizing the sub-arcsecond angular resolution capabilities of the Advanced CCD Imaging Spectrometer (ACIS) in the soft X-ray band (0.3-1.2 keV). These lobes were found to coincide with optical hot spots seen by \textit{Hubble} (\cite{Park2002}), which likely originate from the shock heated gas of the equatorial ring. The hard X-ray emission (1.2-8.0 keV) was found to correlate best with radio emissions observed with the Australia Telescope Compact Array (ATCA) (\cite{Park2002}, \cite{Gaensler1997}, \cite{Ng2013}).

After brightening by more than two orders of magnitude, the soft X-ray flux has leveled off in recent years as the blast wave moves into the lower-density region surrounding the ER (\cite{Frank2016}, \cite{Larsson2019}). In the very-near future, the shocked ER is expected to begin cooling and exhibit a decrease in the flux of soft X-rays. The reverse shock will soon begin heating the heavy elements contained in the SN ejecta, and overtake the ER as the dominant source of X-rays (\cite{Orlando2015},\cite{Orlando2020}). This era will be marked by observable changes in temperature and ionization structure of the remnant, as well as an increase in heavy element abundances. 

In this paper, we analyze the high resolution dispersed spectra from 11 epochs of \textit{Chandra} HETG monitoring observations spanning the time period from 2011-2018. Based on the broadband spectral model fits to these 1$^{st}$-order dispersed spectra, we establish that the metal abundances have not significantly changed during our observation time period, and the temporal evolution of other spectral parameters are generally consistent with the physical picture of the blast wave having entered the lower density region surrounding the inner ring. Information about the observations used in this work is provided in Section \ref{sec:Observations}, while a description of the data analysis and global models is given in Section \ref{sec:Analysis}. A discussion of results can be found in Section \ref{sec:Discussion} and conclusions in Section \ref{sec:Conclusion}.

\section{Observations and Data Reduction} \label{sec:Observations}
\begin{figure}
  \centering
  \includegraphics[width = 0.85\textwidth]{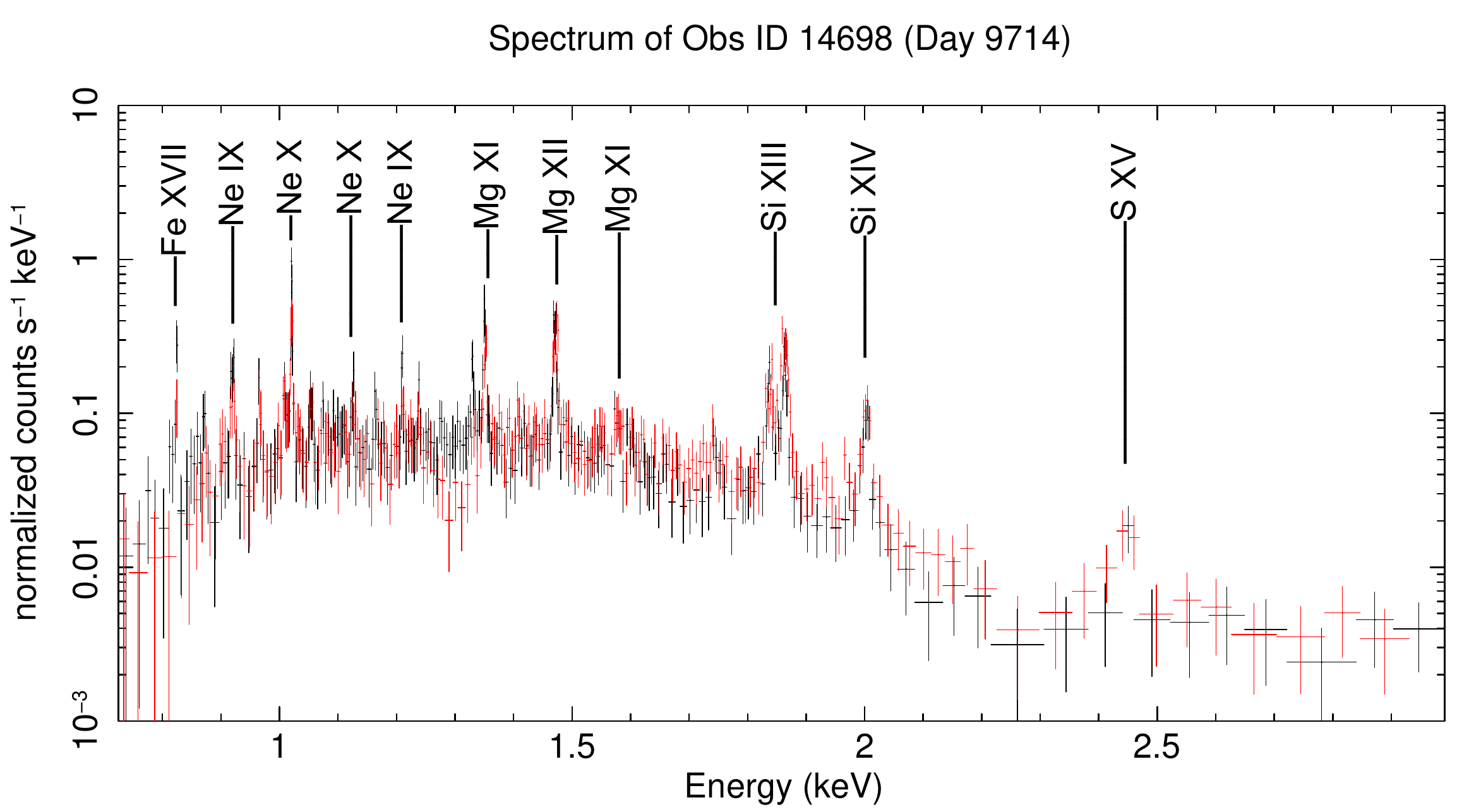}
  \caption{A 1$^{st}$-order ACIS HETG spectrum of SNR 1987A from Obs ID 14698. The MEG-1 and MEG+1 spectra are shown in black and red, respectively. For visualization purposes, data have been binned to 20 counts per bin.}
\label{fig:spectrum}
\end{figure}

The \textit{Chandra} observations used in this work are summarized in Table \ref{table:Observations}. All observations were performed in the ACIS/HETG configuration. Observing parameters such as aim point and subarray size have been altered in order to mitigate the effects of pileup as the remnant brightens, along with inclusion of the diffraction gratings since July 2008 (\cite{Helder2013}). The positive and negative 1$^{st}$-order MEG spectra were extracted following the Science Threads for Grating Spectroscopy in the CIAO 4.12 software package, and response functions were generated with the CALDB 4.9.0 database. Each epoch represents a total exposure of approximately $\sim$50-70ks. A sample spectrum from the observation with the greatest number of counts is shown in Figure \ref{fig:spectrum}. 

The telescope was oriented such that the dispersion direction is aligned with the minor (N-S) axis of the inner ring of SNR 1987A. Because of this orientation, the anticipated spatial-spectral effects produce narrowed spectral lines in the MEG+1 arm and broadened spectral lines in the MEG-1 arm (\cite{Zhekov2005}). The observations are performed with the aimpoint at the center of the chip where contamination of the optical blocking filter is best characterized.

\section{Analysis} \label{sec:Analysis}
The observations used in this work are part of our long-term regular monitoring program of SNR 1987A with \textit{Chandra}. The main goals of this program are to monitor the temporal changes in the X-ray spectrum and morphology of this young SNR based on our semi-annual Chandra ACIS observations of SNR 1987A (e.g. \cite{Burrows2000}, \cite{Park2002}, \cite{Park2004}). As the shock approached the dense ER, the soft X-ray flux dramatically increased by almost two orders of magnitude, resulting in a significant amount of photon pileup in the bare ACIS data. To mitigate this effect, we switched our instrument configuration to insert gratings (either HETG or LETG) in the optical path of the telescope since July 2008. 

Since the inclusion of the HETG in the optical path in July 2008, our Chandra monitoring program has relied on the 0$^{th}$-order imaging and spectral analysis (\cite{Park2011}, \cite{Helder2013}, \cite{Frank2016}). In parallel with these semi-annual monitoring observations, we have also performed several high resolution X-ray spectroscopic studies of SNR 1987A based on the 1$^{st}$-order dispersed spectra taken with deep ($>$200ks) Chandra HETG/LETG observations (e.g. \cite{Zhekov2005}, \cite{Zhekov2006}, \cite{Zhekov2009}, \cite{Ravi2020}). In this work, we apply a similar analysis to the 1$^{st}$-order dispersed spectra taken during our semi-annual monitoring observation to bridge the gap between our two distinct approaches.

To maximize the utility of the high resolution spectroscopy of the 1$^{st}$-order dispersed spectra, we use the unbinned spectra and apply Cash statistics in our spectral analysis (\cite{Cash1979}). Model fitting is restricted to the high signal-to-noise band (0.5-3.0 keV), and all uncertainties represent 1$\sigma$ statistical errors unless otherwise stated. For the broadband spectral model fits, we use a vpshock+vequil model in XSPEC, following our previous approach (e.g. \cite{Frank2016}, \cite{Zhekov2009}). We utilize the non-equilibrium ionization (NEI) version 3.0 and atomic information from AtomDB 3.0.9. Our spectral model and fit results are described in Section \ref{sec:Global Model}.


\vskip 0.5in

\begin{table}
\centering
\begin{tabular}{ccccc}
\hline
Obs ID & Date & \begin{tabular}[c]{@{}c@{}}Exposure \\ (ks)\end{tabular} & \begin{tabular}[c]{@{}c@{}}MEG$\pm$1$^{st}$-order Counts \\ (0.5-3.0 keV)\end{tabular} & \begin{tabular}[c]{@{}c@{}}Age \\ (days)\end{tabular} \\ \hline
12539 & 2011-03-25 & 52.2 & 6356 & 8795 \\ \hline
12540 & 2011-09-21 & 37.5 & \multirow{2}{*}{6645} & \multirow{2}{*}{8975} \\
14344 & 2011-09-22 & 11.5 &  &  \\ \hline
13735 & 2012-03-28 & 42.9 & \multirow{2}{*}{9450} & \multirow{2}{*}{9167} \\
14417 & 2012-04-01 & 26.9 &  &  \\ \hline
14697 & 2013-03-21 & 67.5 & 11420 & 9523 \\ \hline
14698 & 2013-09-28 & 68.5 & 11505 & 9714 \\ \hline
15809 & 2014-03-19 & 70.5 & 10867 & 9886 \\ \hline
17415 & 2014-09-17 & 19.4 & \multirow{2}{*}{11458} & \multirow{2}{*}{10069} \\
15810 & 2014-09-20 & 48.3 &  &  \\ \hline
16756 & 2015-09-17 & 66.6 & 9833 & 10433 \\ \hline
17889 & 2016-09-19 & 26.1 & \multirow{2}{*}{9705} & \multirow{2}{*}{10803} \\
19882 & 2016-09-23 & 41.1 &  &  \\ \hline
20793 & 2017-09-21 & 48.3 & \multirow{2}{*}{6411} & \multirow{2}{*}{11169} \\
19289 & 2017-09-23 & 18.9 &  &  \\ \hline
20277 & 2018-09-15 & 33.8 & \multirow{2}{*}{5913} & \multirow{2}{*}{11527} \\
21844 & 2018-09-16 & 33.8 &  &  \\ \hline
\end{tabular}
\caption{\textit{Chandra} observations utilized in this paper. All observations were performed in the ACIS/HETG configuration. The 0$^{th}$-order image and spectral analysis done by \cite{Frank2016} spans until Obs ID 16756 (day $\sim$10433).}
\label{table:Observations}
\end{table}

\subsection{Broadband Spectral Model and Fit Results} \label{sec:Global Model}
In the literature, it has been firmly established that the observed X-ray spectrum of SNR 1987A can be fitted with a two-component shock model (e.g. \cite{Dewey2008}, \cite{Zhekov2009}, \cite{Frank2016}). Here we adopt the approach taken by \cite{Frank2016}, who performed a comprehensive spectral analysis of the bare-ACIS and 0$^{th}$-order HETG spectra for the epochs partially overlapping with those of this work. Thus, we fit our 1$^{st}$-order dispersed spectra with a \textit{vequil + vpshock} model (see \cite{Frank2016} for the detailed description). In our spectral analysis, we fixed abundances for He and C at the values measured by \cite{Lundqvist1996} for the circumstellar ring of SNR 1987A (since emission from shocked gas in the circumstellar ring dominates the X-ray spectrum), while Ar, Ca and Ni are held at values that are representative of the Large Magellanic Cloud (\cite{Russell1992}). We also fixed the foreground hydrogen column at N$_{H}$ =  2.35 $\times$ 10$^{21}$ cm$^{-2}$ (\cite{Park2006}). Due to the lack of low-energy counts, the abundances of N and O are held constant in this analysis at the best-fit values found by \cite{Zhekov2009}. Parameters that are allowed to vary include temperatures, ionization lifetimes ($\tau$ = n$_{e}t$, where n$_{e}$ is electron density and $t$ is time), normalizations, and abundances for Ne, Mg, Si, S, and Fe. All other abundances are fixed at LMC values (hereafter, abundances are with respect to solar (\cite{Anders1989}) in order to remain consistent with previous works in this observing campaign).



\begin{figure}
  \centering
  \includegraphics[width = 0.9\textwidth]{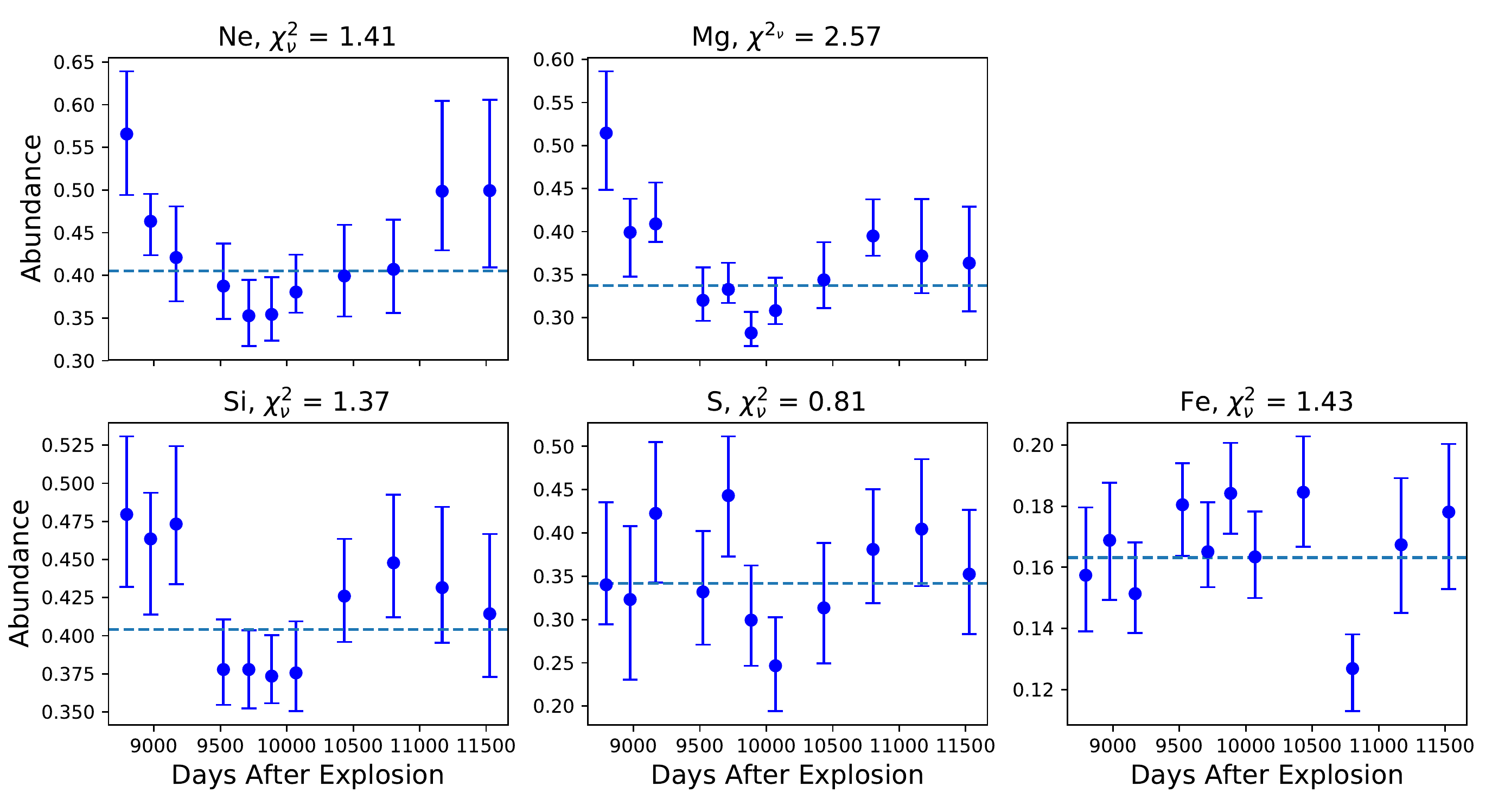}
  \caption{The best-fit abundances of Ne, Mg, Si, S, and Fe over the last 8 years of monitoring observations. Abundances are with respect to solar (\cite{Anders1989}). The weighted mean values are indicated with dashed lines, and the $\chi^{2}_{\nu}$ values corresponding to the weighted means are shown in each panel. These results are also summarized in Table \ref{table:parameters}.}
\label{fig:abundances}
\end{figure}

\begin{figure}
  \centering
  \includegraphics[width = 0.9\textwidth]{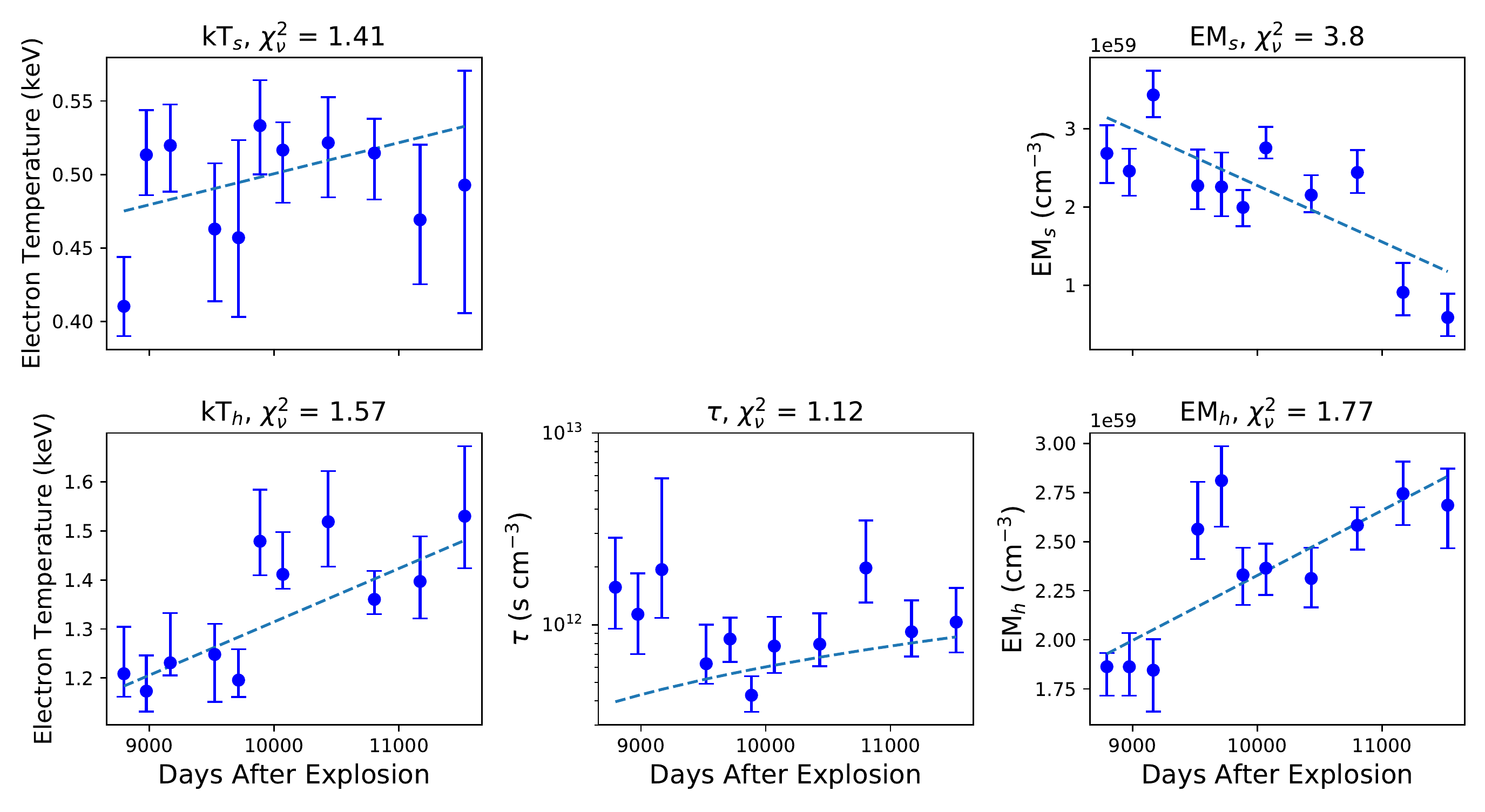}
  \caption{The trends for various best-fit spectral model parameters. The best-fit linear regressions are shown with dashed lines, and the corresponding $\chi^{2}_{\nu}$ values are given in each panel. These results are also summarized in Table \ref{table:parameters}. }
\label{fig:other_parameters}
\end{figure}

We individually fit the broadband X-ray spectra extracted from all 11 epochs to the two-component shock models described in Section \ref{sec:Global Model}. The best fit abundances over this time are shown in Figure \ref{fig:abundances}, while the best fit values for electron temperatures, ionization timescales, and volume emission measures are shown in Figure \ref{fig:other_parameters}. Volume emission measure (EM) is defined as

\begin{equation}
EM = \int n_{e} n_{i} dV \quad ,
\label{eq:Emission Measure}
\end{equation}

\noindent where n$_{e}$ is the electron density and n$_{i}$ is the ion density, integrated over the volume of the remnant. In order to test the time-variability of these best-fit parameter values over the time period explored here, we calculate a weighted mean for each parameter and perform a $\chi^{2}$ analysis to these fits. We conclude that the elemental abundances are well-fit by a constant value over this time interval (Ne = 0.40$\pm$0.02; Mg = 0.34$\pm$0.02; Si = 0.40$\pm$0.01; S = 0.34$\pm$0.02; Fe = 0.16$\pm$0.01; 1$\sigma$ errors). A summary of these parameters can be seen in Table \ref{table:parameters}. Other parameters, such as temperature and volume emission measure, appear to be changing over time as shown in Figure \ref{fig:other_parameters}.




\section{Discussion} \label{sec:Discussion}

Our fitted elemental abundance values did not change significantly over the previous 8 years. This is consistent with the evolutionary path predicted by \cite{Orlando2020}, in which the equatorial ring continues to be the dominant source of X-ray emission over the course of this period. We note that our estimated abundances for Ne, Mg, Si, and S determined through this analysis are on average 23$\%$ higher than those found in the deep 2007 ACIS/HETG data (\cite{Zhekov2009}). This is likely due to the newer AtomDB version used in this work, which produces higher abundance values than identical analyses performed with previous versions. In Table \ref{table:comparisons} we provide comparisons of our best-fit model abundances to those found by studies of the LMC and SNR 1987A in optical wavelengths (\cite{Mattila2010, Russell1992}). Also shown are the solar abundance values of \cite{Anders1989}, which are commonly used in X-ray studies of SNR 1987A, and the more recent solar abundances found by \cite{Asplund2009}. While the abundance of N and O appear to be significantly lower for studies derived from X-ray observations of the SNR, the abundance of Ne and Fe are found to be in good agreement with the results of the optical survey presented in \cite{Mattila2010}.

Although the broadband spectral model fits presented in this work were found by freezing the He abundance to that of \cite{Lundqvist1996}, the findings of \cite{Mattila2010} indicate that the true He abundance of the ER may be lower than previously estimated. We briefly explored the effects of utilizing the He abundance determined by \cite{Mattila2010} (which represents a $\sim$32$\%$ decrease), and found that the resulting best-fit abundance values were almost uniformly $\approx$16$\%$ lower. The emission measure for both the hard (EM$_{h}$) and soft (EM$_{s}$) components also decreased by $\approx$13$\%$, but temperatures and ionization lifetime remained largely unaffected by this change.


Also of note is that our estimated electron temperatures of the hot shock component, shown in Figure \ref{fig:other_parameters}, are markedly lower in all epochs than those in previous works. Our analysis indicates a mean value of $\sim$1.3 keV for the hot component, although it appears to be trending upward somewhat through the epochs explored here. For observations prior to 2015, \cite{Frank2016} cites $\sim$1.8 keV from their analysis of the 0$^{th}$-order image and \cite{Zhekov2009} cites 2.43 keV for the deep ($>$200ks) 2007 HETG observation, which could indicate a systematic difference in temperature measurements found by the two techniques. Determination of whether or not this trend is truly physical will require observations with significantly improved photon statistics. Such an analysis is outside the scope of this paper, and will be presented in a separate work based on our deep HETG data taken in March 2018 (\cite{Ravi2020}, in prep.).

The ratio of emission measures between the hard and soft components varies substantially in our analysis, with a value of EM$_{s}$/EM$_{h}$ $\approx$ 1.5 around day 9000, and gradually shifting to a value of EM$_{s}$/EM$_{h}$ $\approx$ 0.5 by day 11500. In comparison, the analysis of \cite{Zhekov2009} finds that EM$_{s}$/EM$_{h}$ $\approx$ 2 at day 7500. Previous works showed that the soft X-ray light curve has appeared to flatten out since day 7500, while the hard light curve continues to gradually rise (\cite{Park2011},\cite{Helder2013}, \cite{Frank2016}). This is consistent with the evolutionary viewpoint that the supernova blast wave has entered the less-dense region surrounding the equatorial ring, and has begun heating an increasingly large volume of it to X-ray emitting temperatures.


\section{Conclusion} \label{sec:Conclusion}
We have utilized the dispersed spectral data from 11 epochs of \textit{Chandra} observations to analyze various spectral fit parameters over an eight-year span, and compared them to a number of works that utilized similar observations in both X-rays and optical wavelengths. By adopting a two-shock model that is utilized in previous works, we have demonstrated that various observed chemical abundances have remained constant within statistical uncertainties during this period of time (95$\%$ confidence). This is in agreement with previous numerical simulations of the evolution of SNR 1987A, and provides a baseline for detecting the changes that the remnant is expected to undergo in the coming years.

We have explored the viability of using dispersed spectral data from the semi-annual monitoring observations to follow the evolution of spectral characteristics over time, which can help guide future observation requirements in order to adequately meet science goals. These results can be used in conjunction with the less-frequent deep observations to provide a more thorough picture of SNR 1987A as it enters the next phase of its evolution.

\clearpage
\section*{Acknowledgments}
We would like to thank Svet Zhekov for providing critical guidance and feedback throughout the data reduction and model fitting process. This work was supported by NASA grant NNX16AO90H, as well as SAO grants GO1-12070X,  SAO GO2-13064X, SAO GO3-14058X, SAO GO4-15056X, SAO G05-16054X, SAO GO6-17058X, SAO GO7-18060X, and GO8-19043A. Sangwook Park and Aravind Ravi have been supported in part by the Smithsonian Astrophysical Observatory (SAO) through the Chandra grants GO7-18060A, GO8-19043B, and GO9-20051B.
\clearpage

\begin{table}
\centering
\begin{tabular}{lcccc}
\hline
Parameter & \begin{tabular}[c]{@{}c@{}}Best-fit \\ weighted mean\end{tabular} & \begin{tabular}[c]{@{}c@{}}$\chi^{2}_{\nu,constant}$\\ (10 DoF)\end{tabular} & \begin{tabular}[c]{@{}c@{}}$\chi^{2}_{\nu,linear}$\\ (9 DoF)\end{tabular} & \begin{tabular}[c]{@{}c@{}}MEG 2007$^{a}$\\ $[$90$\%$ CI]\end{tabular} \\ \hline
N$_{H}$ (10$^{21}$ cm$^{-2}$) & 2.35$^{b}$ & . . . & . . . & 1.30 [1.18-1.46] \\
He & 2.57$^{c}$ & . . . & . . . & 2.57$^{c}$ \\
C & 0.09$^{c}$ & . . . & . . . & 0.09$^{c}$ \\
N & 0.56$^{a}$ & . . . & . . . & 0.56 [0.50-0.65] \\
O & 0.081$^{a}$ & . . . & . . . & 0.081 [0.074-0.092] \\
Ne & 0.40 $\pm$ 0.02 & 1.41 & 1.47 & 0.29 [0.27-0.31] \\
Mg & 0.34 $\pm$ 0.02 & 2.57 & 2.82 & 0.28 [0.26-0.29] \\
Si & 0.40 $\pm$ 0.01 & 1.37 & 1.51 & 0.33 [0.32-0.35] \\
S & 0.34 $\pm$ 0.02 & 0.81 & 0.90 & 0.30 [0.24-0.36] \\
Ar & 0.537$^{d}$ & . . . & . . . & 0.537$^{d}$ \\
Ca & 0.339$^{d}$ & . . . & . . . & 0.339$^{d}$ \\
Fe & 0.16 $\pm$ 0.01 & 1.43 & 1.53 & 0.19 [0.19-0.21] \\
Ni & 0.618$^{d}$ & . . . & . . . & 0.618$^{d}$ \\
$\tau$ (10$^{11}$ cm$^{-3}$ s) & 6.0 $\pm$ 2.7 & 1.13 & 1.12 & 2.23 [1.81-2.76] \\ \hline
\multicolumn{5}{l}{$^{a}$\footnotesize{\cite{Zhekov2009}}, which utilizes AtomDB 2.0} \\
\multicolumn{5}{l}{$^{b}$\footnotesize{\cite{Park2006}}} \\
\multicolumn{5}{l}{$^{c}$\footnotesize{\cite{Lundqvist1996}}} \\
\multicolumn{5}{l}{$^{d}$\footnotesize{\cite{Russell1992}}}
\end{tabular}
\caption{A summary of broadband spectral model parameters, as well as best-fit constant values. The corresponding $\chi^{2}$ values for both constant and linear fits are also shown. For a p-value of 0.05, $\chi^{2}_{critical}$=18.3 (16.9) for 10 (9) degrees of freedom.}
\label{table:parameters}
\end{table}

\begin{table}
\centering
\begin{tabular}{ccccccc}
\hline
Element & \begin{tabular}[c]{@{}c@{}}Solar \\ (AG89)\end{tabular} & \begin{tabular}[c]{@{}c@{}}Solar\\ (A09)\end{tabular} & \begin{tabular}[c]{@{}c@{}}LMC \\ (RD92)\end{tabular} & \begin{tabular}[c]{@{}c@{}}SNR 1987A\\ (optical) (M10)\end{tabular} & \begin{tabular}[c]{@{}c@{}}SNR 1987A\\ (X-ray) (Z09)\end{tabular} & This Work \\ \hline
He & 10.93(0.01) & 10.98(0.01) & 10.94(0.03) & 11.23$^{+0.13}_{-0.19}$ & . . . & . . . \\
C & 8.56(0.04) & 8.43(0.05) & 8.04(0.18) & . . . & . . . & . . . \\
N & 8.05(0.04) & 7.83(0.05) & 7.14(0.15) & 8.44$^{+0.15}_{-0.22}$ & 7.81(0.06) & . . . \\
O & 8.93(0.04) & 8.69(0.05) & 8.35(0.06) & 8.27$^{+0.08}_{-0.10}$ & 7.85(0.05) & . . . \\
Ne & 8.09(0.10) & 7.93(0.10) & 7.61(0.05) & 7.93 & 7.55(0.03) & 7.69(0.04) \\
Mg & 7.58(0.05) & 7.60(0.04) & 7.47(0.13) & . . . & 7.03(0.02) & 7.11(0.05) \\
Si & 7.55(0.05) & 7.51(0.03) & . . . & . . . & 7.07(0.02) & 7.15(0.02) \\
S & 7.21(0.06) & 7.12(0.03) & 6.77(0.13) & 7.12$^{+0.19}_{-0.34}$ & 6.68(0.09) & 6.74(0.05) \\
Ar & 6.56(0.10) & 6.40(0.10) & 6.29(0.25) & 6.23 & . . . & . . . \\
Ca & 6.36(0.02) & 6.34(0.04) & 5.89(0.16) & 6.51 & . . . & . . . \\
Fe & 7.67(0.03) & 7.50(0.04) & 7.23(0.14) & 6.98$^{+0.19}_{-0.34}$ & 6.97(0.02) & 6.87(0.05) \\ \hline
\end{tabular}
\caption{A comparison of elemental abundances in units of 12+log[\textit{n}(X)/\textit{n}(H)] for a number of works, with errors given in parenthesis where available. AG89: \citep{Anders1989}. A09: \citep{Asplund2009}. RD92: \citep{Russell1992}. M10: \citep{Mattila2010}. Z09: \citep{Zhekov2009}.}
\label{table:comparisons}
\end{table}

\newpage\
\newpage

\bibliography{bibliography_database}{} 

\end{document}